# 3D Unconventional Superconductivity in Bulk LaO

*Zhifan Wang,[1,#] Jingkai Bi,[2,#] Jiayuan Zhang,[3,#] Wenmin Li,[2] Yuxuan Liu,[4] Dao-Xin Yao,[3,*] Zheng Deng,[5,*] Changqing Jin,[5] Yifeng Han,[1,*] Man-Rong Li,[1,*]*

[1]School of Chemistry and Chemical Engineering, Hainan University, Haikou 570228, P. R. China

[2]Institute of Quantum Materials and Physics, Henan Academy of Sciences, Zhengzhou 450046, P. R. China

[3]State Key Laboratory of Optoelectronic Materials and Technologies, Guangdong Provincial Key Laboratory of Magnetoelectric Physics and Devices, School of Physics, Sun Yat-Sen University, Guangzhou 510275, P. R. China

[4]Key Laboratory of Bioinorganic and Synthetic Chemistry of Ministry of Education, School of Chemistry, Sun Yat-sen University, Guangzhou, 510006, P. R. China

[5]Beijing National Laboratory for Condensed Matter Physics, Institute of Physics, Chinese Academy of Sciences, Beijing 100190, P. R. China

[#]*These authors contributed equally to this work*

Email: yaodaox@mail.sysu.edu.cn (D.-X.Yao); dengzheng@iphy.ac.cn (D. Zheng); Yifenghan@hainanu.edu.cn (Y. Han); limanrong@hainanu.edu.cn (M.-R. Li)

**Abstract**

Lanthanum-based compounds are cornerstones of superconductivity research, yet the La-5$d$ orbitals typically remain empty spectator states far above the Fermi level ($E_F$). While superconductivity has been induced in LaO up to 5.37 K in tensile epitaxy films, the intrinsic ground state of the bulk phase has remained controversial mostly due to synthetic challenges, with early reports suggesting a metallic nature. Here we report the high-pressure and high-temperature synthesis of pure bulk rock-salt LaO and unveil its intrinsic type-II superconductivity with a transition temperature ($T_C$) of ~6 K at ambient pressure. The bulk $T_C$ is further enhanced to 6.9 K in La$_{1-x}$Y$_x$O at $x$ = 0.10, where Y doping leads to lattice contraction (chemical pressing) and a remarkable increase in electron carrier concentration. Strikingly, applying physical pressure further enhances the $T_C$ to a maximum of 12.7 K at 20 GPa, the highest $T_C$ in lanthanum monochalcogenides La$X$ ($X$ = S, Se, Te, and O) to date. This pressure dependence is diametrically opposed to the behavior observed in films, and occurs despite a pressure-induced reduction in the density of states at $E_F$ - a trend that sharply contradicts the conventional phonon-mediated BCS mechanism. Our first-principles calculations reveal that compressive strain modifies the crystal field splitting to enhance La-5$d$/O-2$p$ hybridization, fostering a three-dimensional multi-pocket Fermi surface favorable for spin/orbital fluctuation-mediated pairing. This work clarifies the intrinsic superconductivity of bulk LaO and provides a foundation for designing new rare-earth-based superconductors with higher $T_C$.

**1. Introduction**

Lanthanum-based compounds constitute one of the most fertile grounds for discovering high-temperature superconductivity, exemplified by the pioneering cuprates [1,2], the recently discovered nickelates [3-5], and iron-based superconductors [6]. In the vast majority of these systems, however, the lanthanum ion adopts a stable trivalent state (La$^{3+}$) [7]. In this configuration, the 5$d$ and 5$s$ orbitals remain empty and energetically remote from the $E_F$, relegating La to the role of a passive structural spacer or charge reservoir. A rare and intriguing departure from this paradigm is found in the lanthanum monochalcogenides La$X$ ($X$ = S, Se, Te, O), where lanthanum adopts a divalent state (La$^{2+}$) [8-10]. Here, the typically dormant 5$d$ electrons become active, directly populating the conduction bands, and offering a unique frontier to explore unconventional quantum phases driven by 5$d$-orbital physics. Among this family [11,15], lanthanum monoxide (LaO) stands out as a particularly enigmatic candidate. While thermodynamic considerations suggest it should crystallize in a simple rock-salt structure, its synthesis requires stringent high-pressure and high-temperature (HPHT) conditions. Early studies on bulk LaO two decades ago reported a metallic ground state, leaving the question of its superconductivity dormant. Interest was reignited recently when superconductivity up to 5.37 K was observed in epitaxial LaO thin films [12-16]. However, interpretation of these film results is complicated by substrate-induced tetragonal distortions, where tensile strain was seemingly required to stabilize the superconducting phase. This has led to a dichotomy in the field: Is superconductivity intrinsic to the ideal rock-salt LaO lattice, or is it an artifact of interfacial strain? Furthermore, while initial theoretical works proposed a conventional phonon-mediated mechanism [15-17], they struggled to reconcile the relationship between lattice volume ($V$) and superconducting temperature ($T_C$) across different experimental regimes.

Resolving these fundamental questions requires accessing the pristine bulk phase, free from substrate clamping. Unlike thin films, where lattice parameters are pinned, bulk LaO allows for the continuous tuning of the lattice via hydrostatic pressure and chemical doping, providing a clean testbed for the pairing mechanism. Specifically, clarifying how $T_C$ responds to isotropic compression is critical; for a conventional phonon-mediated superconductor, pressure-induced broadening of bands typically suppresses the density of states (DOS) and $T_C$ [17]. Conversely, an enhancement of $T_C$ under pressure-analogous to the behavior seen in cuprates and iron-based systems [18,19] would signal the presence of alternative pairing interactions. Lavroff et al. showed in LaB$_6$ that the localization of Fermi-level states is crucial for electron–phonon coupling (EPC): La 4$f$-B π hybridization suppresses EPC, yielding $T_C$=0.45 K [20]. Zhai et al. found in transition-metal diborides that anisotropic strain selectively tunes metal–boron hybridization and bond covalency, thereby modulating phonon-mediated superconductivity [21]. These studies offer a useful framework for understanding how orbital hybridization and state localization control superconductivity. In this study, we revisit the bulk limit of LaO. By mastering the HPHT synthesis of stoichiometric samples, we not only confirm its intrinsic type-II superconductivity but also unveil a pressure-tunable phase diagram that defies conventional BCS expectations.

## 2. Experimental and Methods

**HPHT synthesis and phase stability**
La$_2$O$_3$ (99.999%, Macklin) and Y$_2$O$_3$ (99.999%, Macklin) powders were annealed at 1273 K in air before transferred into an argon-filled glovebox. La metal (99.7% (metal), mesh 40, Alfa Aesar) was received packed under Ar and opened in a glovebox. The phase analysis of La metal was performed by X-ray diffraction (XRD) in a sealed holder. La was confirmed to be the hexagonal $P6_3/mmc$ phase. La$_{1-x}$Y$_x$O ($x$ = 0, 0.05, 0.10) were synthesized by HPHT method in our 420-type six-sided top press [22]. The raw materials La$_2$O$_3$, La and Y$_2$O$_3$ with certain stoichiometric ratios were fully mixed in an agate mortar in Ar-filled glovebox. Slightly excessive (~3% by weight/mol) La was applied to account for the possible surface oxidation of La metal. Afterward, the mixture was pressed into blocks with a diameter and height of 0.55 and 0.70 cm, respectively, under 5 MPa in a glovebox. Subsequently, the block was assembled into a synthetic block and placed in a six-sided top press. For all HPHT experiments, the pressure was increased to 5 GPa and then the assembly was heated at 1573 K for 2-4 h before quenching to room temperature. The pressure was then slowly decompressed to ambient. The recovered capsules were opened, and the La$_{1-}$ $_x$Y$_x$O ($x$ = 0, 0.05, 0.10) sample pellets were cleaned minimally by breaking or scraping off remaining large pieces of the capsule to remove residue and minimize impurity. Samples were then stored in an Ar-atmosphere glovebox. The synthesized bulk polycrystalline La$_{1-x}$Y$_x$O exhibited golden-yellow color with metallic luster. The phase purity was characterized by powder XRD and **s**ynchrotron powder X-ray diffraction (SPXD). It is noteworthy that the polycrystalline La$_{1-x}$Y$_x$O undergoes gradual degradation when exposed to air. Comparative XRD analyses under different storage conditions revealed that samples stored in Ar atmosphere maintained unchanged, whereas those exposed to air developed prominent La(OH)$_3$ peaks in XRD patterns. Consequently, all samples were stored in our Ar-filled glovebox to preserve their structural integrity.

**Phase and structural analyses**

XRD patterns were obtained using a powder X-ray diffractometer (MiniFlex 600, Rigaku, Japan) equipped with Cu Kα tube (40 kV and 15 mA). The surfaces of the as-made sample pellets were polished smooth and mounted with non-peak modeling clay for XRD data collection to assess phase purity. SPXD data were acquired at the beamline BL14B ($\lambda$ = 0.6886 Å) of the Shanghai Synchrotron Radiation Facility (SSRF) and beamline BL44B2 ($\lambda$ = 0.406586 Å) of SPring-8, Japan. Rietveld refinements of the SPXD data were performed using Topas-Academic *V*6 software [23].

**Magnetic and electrotransport measurements**

Magnetic properties were measured using a Physical Property Measurement System (PPMS-9 T, Quantum Design). The temperature-dependent zero-field cooling (ZFC) and field cooling (FC) magnetic susceptibility data were collected in 2-300 K under an applied magnetic field of 0.003 T. Isothermal magnetization curves were recorded at various temperatures in magnetic fields ranging from −1.5 to 1.5 T. The high-pressure electronic transport properties of La$_{1-x}$Y$_x$O were measured through four-probe electrical conductivity methods in diamond anvil cells (DACs) made of CuBe alloy. The diamond culet is 300 μm in diameter. Au wires with diameter of 18 μm were used as electrodes. A T301 stainless steel gasket was compressed from thickness of 250 to 40 μm, and a hole of 150 μm in diameter was drilled by laser. Cubic BN as an insulating layer was pressed into this hole. A small central hole with diameter of 100 μm was further drilled to serve as the sample chamber, where NaCl fine powder serves as a pressure transmitting medium and a piece of compressed La$_{1-x}$Y$_x$O powder sample with dimensions of 90 × 90 × 20 μm was loaded. A ruby ball was loaded simultaneously as a pressure indicator. The assembled DACs were placed inside a cryogenicsystem with automatic temperature control.

**DFT calculations**

First-principles calculations are performed by using density functional theory (DFT) as implemented in the Vienna *ab initio* simulation package (VASP) [24,25] with the projector augmented wave method [26] and the exchange-correlation functional within the generalized gradient approximation [27]. For the geometric structure relaxation, the total energy and the force on each atom are converged within 10$^{-7}$ eV and less than $10^{-3}$ eV/Å, respectively. A plane-wave cutoff energy of 600 eV is adopted. The Brillouin zone (BZ) is sampled with a 11 × 11 × 11 $k$-point mesh for bulk LaO. The Fermi surface is obtained by interpolating the Hamiltonian on the basis of maximally localized Wannier functions using the Wannier90 package [28,29], where the electronic structure around the Fermi level ($E_F$) is described by 15 Wannier functions (three $p$ states for each O atom and five $d$ and seven $f$ states for the La atom).

Lattice dynamics and electron-phonon calculations are performed by using density functional perturbation theory as implemented in the QUANTUM ESPRESSO package [30,31], with the norm-conserving pseudopotentials, an energy cutoff of 80 Ry, and a sampling of the electronic (vibrational) BZ by 12 × 12 × 12 (6 × 6 × 6) meshes. The electron-phonon matrix elements are calculated using a dense 24 × 24 × 24 $k$-point mesh. The superconducting critical temperature is evaluated through the Allen-Dynes modified McMillan formula [32] with a typical Coulomb pseudopotential $\mu^* = 0.13$ [17].

## 3. Results and Discussion

**Crystal structure**

The as-made bulk LaO crystallizes in the characteristic NaCl-type structure (*Fm-3m*) [33], which is consistent with other reported rare earth monoxides (*RE*Os [34-39], **Figure 1a**). Rietveld refinements of room-temperature SPXD data yielded the lattice parameter of $a = 5.1451(2)$ Å, which is smaller than the previously reported dimensions in LaO thin films (5.351 Å), and consistent with reported one in bulk (5.144 Å). The ionic radius of $La^{2+}$ is estimated to be 1.173 Å in our bulk phase, which is very comparable with the reported value of six-coordinate $La^{2+}$ (1.25 Å) [40]. Specifically, epitaxial LaO films on $YAlO_3$ substrates adopt a reduced lattice parameter (cube root of unit cell) of 5.202 Å, while those grown on $LaAlO_3$ and $LaSrAlO_4$ substrates show expanded lattice parameters of 5.245 and 5.298 Å, respectively, under tensile strain [15]. The significant variations in these lattice parameters mainly originate from the interfacial lattice mismatch between films and substrates, with their lattice mismatch degrees being 38.6%, 91.2% and 93.1%, respectively. This effect will induce different degrees of lattice strain during the epitaxial growth process. Upon Y doping in $La_{1-x}Y_xO$, XRD peaks shift to higher diffraction angles with increasing Y content ($x$) up to 0.10 (**Figure 1b**), indicating a decrease in the lattice parameter $a$ [41]. For $x = 0.15$, diffraction peaks corresponding to a secondary phase $LaYO_3$ appear. Refinements confirmed the decreasing trend in the primary phase, yielding lattice parameters $a$ of 5.1451(2), 5.1398(6), and 5.1294(4) Å for $x = 0$, 0.05, and 0.10, respectively (**Figure S1**). These findings suggest that the solubility limit of Y in LaO under the employed synthesis conditions is approximately $x = 0.10$ within the experiment error.

**Magnetism**

The superconducting properties of $La_{1-x}Y_xO$ ($x$ = 0, 0.05, 0.10) were firstly investigated through magnetization measurements under ambient pressure. **Figure 1c** presents the temperature dependence of magnetization, $M(T)$, measured in ZFC and FC modes under an applied field of 30 Oe for $x$ = 0, 0.05, and 0.10. All samples exhibit a distinct diamagnetic transition with an estimated superconducting shielding fraction approaching 100%, confirming bulk superconductivity. The bulk superconductivity is further evidenced by Specific heat (**Figure S2**). The onset transition temperatures ($T_C$(mag)), determined from the $M(T)$ curves, are approximately 5.58, 5.95, and 6.38 K for $x$ = 0, 0.05, and 0.10, respectively. Note that La is a superconductor with $T_C$ = 4.9 K, thus the possibility of superconductivity from La can be excluded. Magnetization versus magnetic field, $M(H)$, curves were measured for the parent LaO sample (**Figure 1d**), exhibiting typical type-II superconductivity.

**Electrical conductivity**

To investigate the normal-state and superconducting characteristics, electrical resistivity measurements were performed on $La_{1-x}Y_xO$ ($x$ = 0, 0.05, 0.10). As shown in Supplementary **Figure S3**, the normal-state temperature dependence of resistivity, $\rho(T)$, displays typical metallic behavior for all compositions. **Figure 1e** provides an enlarged view of the superconducting transition region under ZFC. These $T_C$ are consistent with magnetization measurements. Notably, the $T_C$ of our bulk LaO (6.05 K) is approximately 1 K higher than the previously reported value of ~5 K in LaO thin films. Supplementary **Figure S4** shows the $\rho(T)$ curves under various applied magnetic fields for representative samples. The criterion of 90% $\rho_n$ ($\rho_n$: normal-state resistivity) is used to estimate upper critical field ($H_{C2}$). The temperature dependences of $H_{C2}$ of Y = 0, 0.05 and 0.10 are shown in **Figure 1f**. Using the Ginzburg–Landau (GL) formula, $H_{C2}(T) = H_{C2}(0) \times (T_{C0}^2-T^2)/(T_{C0}^2+T)$, where $T_{C0}$ is the $T_C$ under zero magnetic field, and $H_{C2}(0)$ is $H_{C2}$ at zero temperature, the largest $H_{C2}(0)$ ($\approx$ 3.85 T) is obtained for Y = 0.10. The coherence length at zero temperature ($\xi(0)$) can be estimated using $H_{C2}(0) = \Phi_0/2\pi\xi(0)^2$, where $\Phi_0 = h/2e$ is the magnetic flux quantum. We obtain $\xi(0)$ = 93 Å for Y = 0.10, and it is comparable with nickelates [42].

**Chemical and physical pressure-enhanced conductivity**

Substituting La with smaller Y ions induces chemical pressure via lattice contraction [43,44]. Unlike the positive dependence of $T_C$ on the cell expansion in LaO film, the chemical pressure enhances $T_C$ in bulk LaO upon cell contraction. Hall effect measurements performed at 10 K elucidate the normal-state carrier concentration of LaO and $La_{0.9}Y_{0.1}O$. As shown in **Figure S5**, the calculated electron carrier concentration ($n_e$) increases from 2.5 × $10^{22}$ cm$^{-3}$ in LaO to 3.3 × $10^{22}$ cm$^{-3}$ in $La_{0.9}Y_{0.1}O$, indicating that Y doping effectively enhances the electron density in this system. It is consistent with the result that the Y-doped sample has the higher conductivity. The close correlation between carrier concentration and $T_C$ has been studied extensively [45]. Here the positive correlation indicates the possibility of further optimizing $T_C$ with extra carrier concentration.

*In situ* XRD and electrical resistance measurements were performed on LaO up to 53 GPa as shown in **Figure 2a**. No phase transition was observed during the compression as shown in **Figure S6**. The resulting pressure dependence of the superconducting transition temperature, $T_C(P)$, reveals several distinct regimes. Initially, $T_C$ rises sharply from its ambient pressure value of ~6 to 11.8 K by 2.5 GPa. This rapid enhancement is followed by a more gradual increase, with $T_C$ reaching a maximum of 12.7 K around 20.1 GPa. Between approximately 20.1 and 25.3 GPa, $T_C$ exhibits a plateau before gradually declining upon further pressure increase beyond 25.3 GPa. This evolution traces a distinct dome-shaped $T_C(P)$ phase diagram (**Figure 2b**), suggesting the influence of competing pressure-induced effects on superconductivity. Such dome-shaped pressure-dependence reminiscences pressure effects on numerous unconventional superconductors, including iron-based superconductors with multi-orbital multi-band nature. For LaO, the possible band splitting, which is indicated by our DFT calculations in following section, will be clarified experimentally in a separated work in the future.

To clarify the apparent discrepancy between bulk and thin-film LaO, it is instructive to compare their lattice tuning mechanisms. In epitaxial LaO films, superconductivity is stabilized only under tensile strain imposed by the substrate, which introduces anisotropic lattice distortion and reduces the cubic symmetry. In contrast, bulk LaO allows isotropic lattice tuning via chemical substitution and hydrostatic pressure while preserving the rock-salt symmetry, and both routes consistently enhance $T_C$. Notably, the enhancement of $T_C$ under compression in bulk LaO is opposite to the strain dependence reported in thin films, indicating that strain and pressure act through distinct microscopic mechanisms. While anisotropic strain mainly modifies crystal-field anisotropy and band degeneracy, isotropic compression enhances La-5$d$–O-2$p$ hybridization and interorbital hopping, promoting superconductivity in a three-dimensional electronic environment. This contrast highlights bulk LaO as a platform to access the intrinsic superconducting state free from interfacial and substrate-clamping effects.

**Electronic Structure Evolution and Superconducting Mechanism under Compressive Strain**

To understand the superconducting mechanism of bulk LaO, we calculated its electronic properties and superconducting critical temperature ($T_C$) under isotropic compressive strain (uniform compression in *a*, *b*, *c* axes), focusing on orbital hybridization, crystal field splitting, and Fermi surface topology. As shown in **Figure 3**, the calculated band structures and DOS of bulk LaO under unstrained and 8% compressive strain reveal a metallic character, with one band crossing $E_F$. The electronic states near $E_F$ are mainly derived from La 5$d$ orbitals, with minor contributions from La 4*f* orbitals, while O 2*p* orbitals contribute negligibly. Under 8% compressive strain, the DOS at $E_F$ decreases by approximately 17% (from 0.98 to 0.81 states/(eV·f.u.)), suggesting that compression reduces the number of available conducting states. The strain-dependent evolution of the band structures and DOS (**Figure S7-8**) further confirms that compressive strain strongly regulates the electronic structure. With increasing strain, the DOS near $E_F$ continuously decreases. According to conventional BCS theory, such a reduction is unfavorable for Cooper pair formation, and $T_C$ is expected to decrease accordingly. Indeed, the calculated $T_C$ decreases with strain (**Figure S9**), and this trend is also consistent with the theoretical $T_C$ calculation trend reported [17] by Sun et al. However, experimental measurements under hydrostatic pressure show the opposite trend. It is hard to be explained by the phonon-mediated BCS mechanism, but rather likely originates from an unconventional pairing mechanism. Similarly, previous first-principles studies on compressed lanthanum-based compound LaBi also revealed that, the electron-phonon coupling mechanism based on weak BCS theory fails to account for the experimentally observed superconductivity (calculated $T_C$ is close to 0 K, while the experimental $T_C$ reaches up to ~8 K), which also implies the existence of alternative superconducting mechanisms [11]. Compared to the unstrained state (**Figure 3a**), the La 5*d* orbitals under 8% compressive strain (**Figure 3b**) exhibit a notable increase in bandwidth, implying enhanced dispersion and spatial delocalization. Specifically, the bottom of the La 5*d* orbitals at high-symmetry points (such as *X*) shifts downward, while the overall band moves upward relative to $E_F$. Meanwhile, the top of the O 2*p* orbitals shifts upward and shows increased dispersion, indicating a significant enhancement of La 5*d*–O 2*p* orbital hybridization. This dual modulation of the band edges—downshift of La 5d and upshift of O 2*p*-is attributed to the synergistic effects of enhanced crystal-field splitting and strengthened orbital hybridization induced by compressive strain.

As illustrated in the insets of **Figure 3**, the Fermi surface of bulk LaO displays a characteristic 3D topology, defined by multiple closed electron-like pockets symmetrically distributed around the Brillouin zone center. This structure signifies strong 3D dispersions and substantial spatial overlap between La-5*d* and O-2*p* orbitals, aligning with the enhanced hybridization observed in the band structure. The multi-pocket nature implies the coexistence of various Fermi scattering channels, offering favorable conditions for the multiband interactions required for unconventional superconductivity. When subjected to an 8% compressive strain, these electron pockets shrink and deform, concurrently enhancing the 3D connectivity between adjacent pockets. This strain-induced reconstruction reflects increased orbital overlap and electronic itinerancy, consistent with the enlarged hopping parameters and broadened bandwidths presented in **Table 1**. Notably, the existence of multiple nesting vectors in the 3D Fermi surface may amplify collective excitations, such as spin or orbital fluctuations, which are widely regarded as essential mediators of unconventional pairing. This suggests that La-based compounds, particularly those centered on 5*d* orbitals, may host a new class of systems driven by unconventional pairing mechanisms.

As summarized in **Table 1**, the tight-binding parameters quantitatively capture the strain-induced splitting of La 5$d$ orbitals, revealing enhanced crystal-field effects in the octahedral environment (**Figure S9**). Moreover, both intra- and inter-atomic hopping parameters increase significantly-*e.g.*, $t_{[1,0]}^{d_{xy}}$ from −0.9288 to −1.2815 eV and $t_{[1,0]}^{d_{x^2-y^2}-d_{p_x}}$ from −1.4972 to −2.2101 eV—indicating strengthened electron itinerancy and orbital fluctuations. Consequently, compressive strain not only enhances electronic kinetic energy and orbital mixing, but may also promote the formation of an unconventional 3D superconducting pairing state through enhanced spin/orbital fluctuation mechanisms. In summary, bulk LaO exhibits unique electronic structure modulation under compressive strain, suggesting that it may be an unconventional 3D superconductor. This finding provides new insights into its superconducting mechanism and warrants further investigation.

The introduction of interfacial strain has been proven to be an effective way to achieve superconductivity at ambient pressure in state-of-the-art epitaxial thin-films, and thus to chemically realize and intercept the metastable superconductive state observed in bulk under physical pressure [44], as exemplified in nickelate superconductors [4,47-52]. In our bulk LaO, the substitution of smaller $Y^{2+}$ for larger $La^{2+}$ leads to volumetric contraction of the unit cell and enhancement of the $T_C$ from 6.05 to 6.9 K at 10% Y-doping. Meanwhile, the pressing of bulk LaO similarly squeezes the lattice and achieves the $T_C$ as high as 12.7 K around 20 GPa. This maximum $T_C$ is more than doubled than those in epitaxial films [15,16], and even higher than the predicted record $T_C$ of 11.11 K to date [17]. These findings suggest that smaller unit cell favors the superconductivity in bulk LaO, which is, however, in contradiction with the discoveries in thin films, where larger volume upon tensile stress favors higher $T_C$ in films [15,16].

Our experimental results are also inconsistent with DFT calculation based on purely phono-mediated Cooper paring, implying unconventional nature of LaO superconductivity. Future theoretical calculations, and/or *in situ* pressure- and temperature-dependent characterizations, will help to understand the rule for enhancing superconductivity in LaO and related systems with exotic electronic structure in rare-earth families [53-56].

## 4. CONCLUSION

We have definitively resolved the long-standing ambiguity regarding the ground state of lanthanum monoxide. Contrary to early reports of metallic behavior, our high-pressure synthesis reveals that stoichiometric bulk LaO is an intrinsic type-II superconductor with $T_C$ of ~ 6 K. This value serves as a baseline that can be dramatically tuned: by applying "lattice compression"-either chemically via Y-substitution or physically via hydrostatic pressure-we achieved a continuous enhancement of $T_C$, culminating in a maximum of 12.7 K at 20 GPa. This is the highest transition temperature recorded in the lanthanum monochalcogenide family to date.

The most striking feature of our findings is the anomalous response of superconductivity to lattice compression, which provides a "smoking gun" evidence against a conventional phonon-mediated mechanism. In epitaxial thin films, superconductivity was believed to rely on lattice expansion (tensile strain). In stark contrast, we find that compression drives the $T_C$ enhancement in the bulk. Crucially, our DFT calculations show that isotropic compression significantly suppresses the DOS at the $E_F$. According to the standard BCS formalism (specifically, the McMillan formula), such a reduction in phase space for pairing should inevitably suppress $T_C$. The experimental observation that $T_C$ doubles while the DOS decreases fundamentally challenges the phonon-mediated scenario, strongly suggesting that the pairing glue in LaO is electronic in origin.

We propose that this unconventional superconductivity arises from the unique proximity of La-$5d$ orbitals to the Fermi level. Unlike most La-based superconductors where La is trivalent, the divalent La in LaO possesses active $5d$ electrons. Our electronic structure analysis reveals that compressive strain induces a "dual modulation": it simultaneously shifts the La-$5d$ band edges downward and the O-$2p$ bands upward. This energetic convergence enhances $p$-$d$ hybridization and delocalization, triggering a reconstruction of the Fermi surface. The resulting three-dimensional topology features multiple electron-like pockets with substantial nesting vectors, creating ideal channels for pairing mediated by spin or orbital fluctuations.

In summary, bulk LaO represents a paradigm shift from "strain-stabilized" to "intrinsic unconventional" superconductivity. The pressure-induced enhancement of $T_C$ -uncoupled from the density of states-mirrors the behavior of high- $T_C$ cuprates and iron-based superconductors, yet it occurs in a remarkably simple rock-salt structure driven by $5d$ electrons. This establishes LaO not merely as a new superconductor, but as a minimalist platform for dissecting the interplay between $5d$-orbital correlations and unconventional pairing, offering a fresh perspective for designing rare-earth-based quantum materials.

**Corresponding Authors**
D.-X. Yao yaodaox@mail.sysu.edu.cn
D. Zheng dengzheng@iphy.ac.cn
Y. Han yifenghan@ hainanu.edu.cn
M.-R. Li  limanrong@hainanu.edu.cn

**ORCID**
Yifeng Han: 0000-0002-7518-978X

Man-Rong Li: 0000-0001-8424-9134
Dao-Xin Yao: 0000-0003-1097-3802

**Supporting Information**
Specific heat measurements, doped cell evolution, electrical resistance under ambient pressure and applied magnetic fields, Hall resistance, high-pressure XRD, and supplementary theoretical calculation diagrams.

**Acknowledgments**

This research was supported by the National Natural Science Foundation of China (22090041, 22575071, 12404007, 12494591, 92165204, 92565303, 22531001), the Guangdong Basic and Applied Basic Research Foundation (Grant No. 2022B1515120014), Guangdong Provincial Quantum Science Strategic Initiative (GDZX2401010), CAS Project for Young Scientists in Basic Research (No. YSBR-030), MOST of China (No. 2022YFA1403900), the High-level Talent Research Start-up Program Funding of Henan Academy of Sciences (241827046, 242027151, and 241827022), the Fundamental Research Fund of Henan Academy of Sciences (240627005), and Beijing National Laboratory for Condensed Matter Physics (2023BNLCMPKF006). The synchrotron radiation experiments were performed at BL44B2 of SPring-8 with the approval of the Japan Synchrotron Radiation Research Institute (JASRI) (Proposal Nos. 2025A1497) and the authors are grateful for the help of the BL14B beam line of SSRF.

**Author contributions**
Y. H., M. L., and Z. W. conceived the research idea and designed the experiments; Z. W. prepared the samples under high pressure; Z. D. and Y. L. conducted the electrical transport and magnetic property measurements at ambient pressure; W. L. and J. B. performed the electrical transport measurements under high pressure; D. Y. and J. Z. performed the DFT calculations and theoretical analysis; all authors participated in the data analysis and agreed to publish the data. The manuscript was written by Z. W. and Y. H; D. Y., Y. H., M. L., Z. D., and W. L. supervised the research.

**Competing interests**
The authors declare that they have no competing interests.

**Figures and Captions**

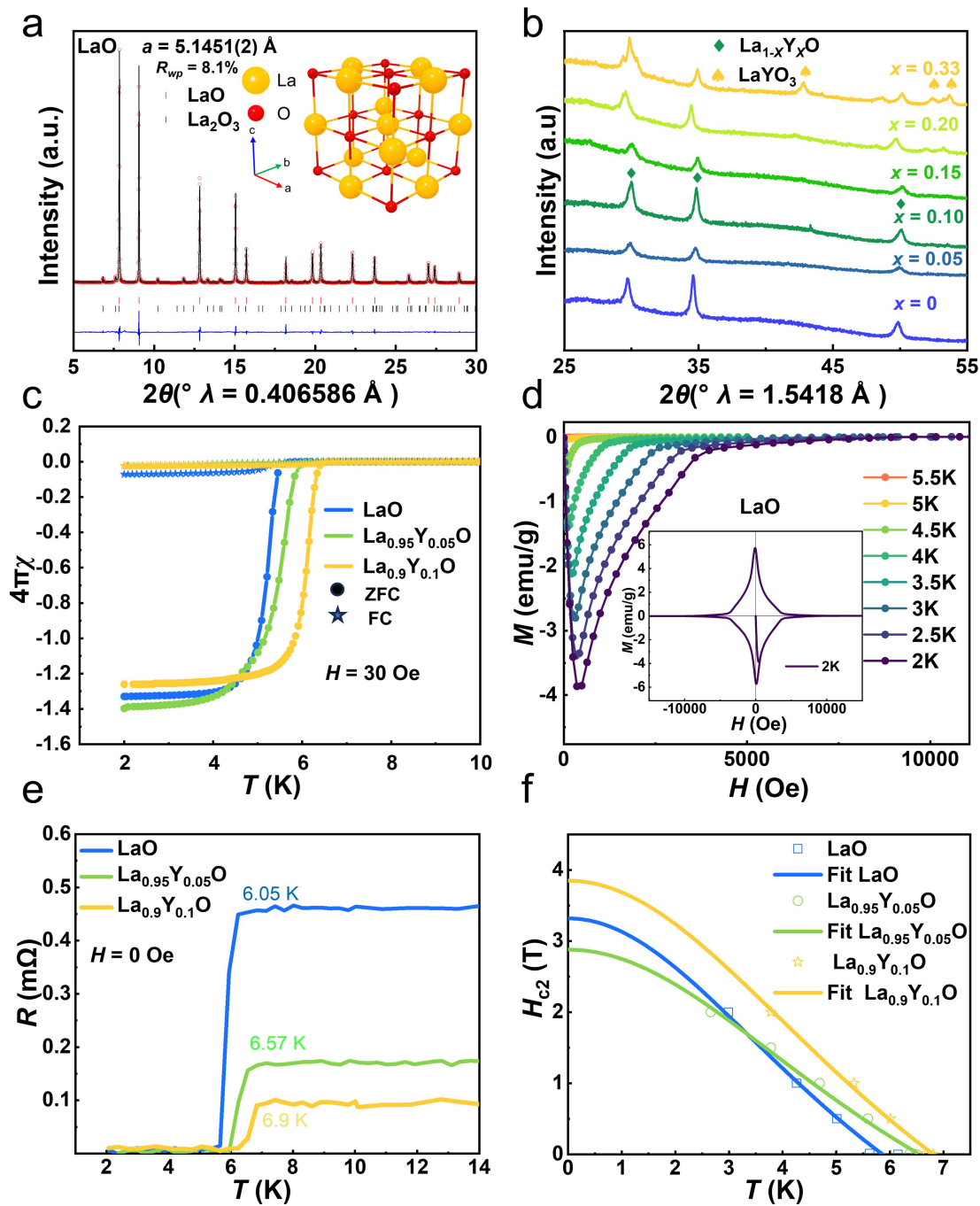

**Figure 1.a.** Rietveld refinements of the SPXD data for LaO. The red circle represents the calculated data, the black line the observed fit, the deep blue line the difference. The upper red and lower black ticks mark the positions of Bragg reflections of LaO ($Fm\bar{3}m$, 97.29%) and $La_2O_3$ ($P\bar{3}m1$, 2.71%), The inset shows the crystal structure of LaO; **b.** XRD patterns of $La_{1-x}Y_xO$ ($x$ = 0, 0.05, 0.10, 0.15, 0.20, 0.33); **c.**

Temperature dependence of the superconducting fraction for $La_{1-x}Y_xO$ ($x$ = 0, 0.05, and 0.10,); **d.** The ($H$) curves of LaO measured below $T_C$ in the field range of 0-1 T, with the inset displaying the $M(H)$ curves at 2 K over an extended field range of -1 to 1 T; **e.** Temperature dependence of resistance for $La_{1-x}Y_xO$ ($x$ = 0, 0.10, 0.15) at ambient pressure; **f.** The temperature dependences of $H_{c2}$ of Y = 0, 0.05 and 0.10.

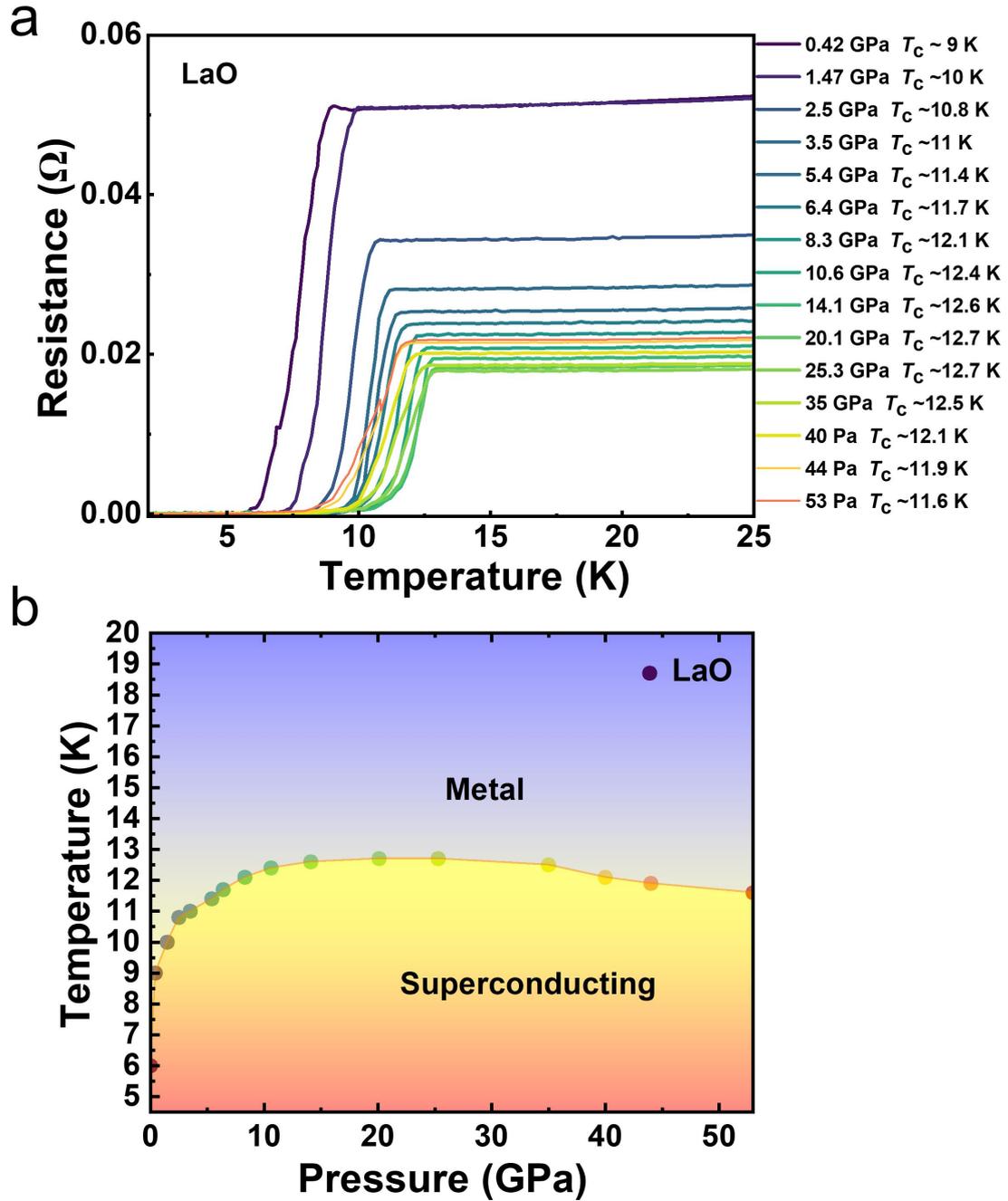

**Figure 2. a.** Temperature dependence of the resistance of LaO at various pressures; **b.** Pressure dependence of the superconducting transition temperature in LaO. (The sample exhibits superconductivity in the orange-yellow region and metallic properties in the blue-white region)

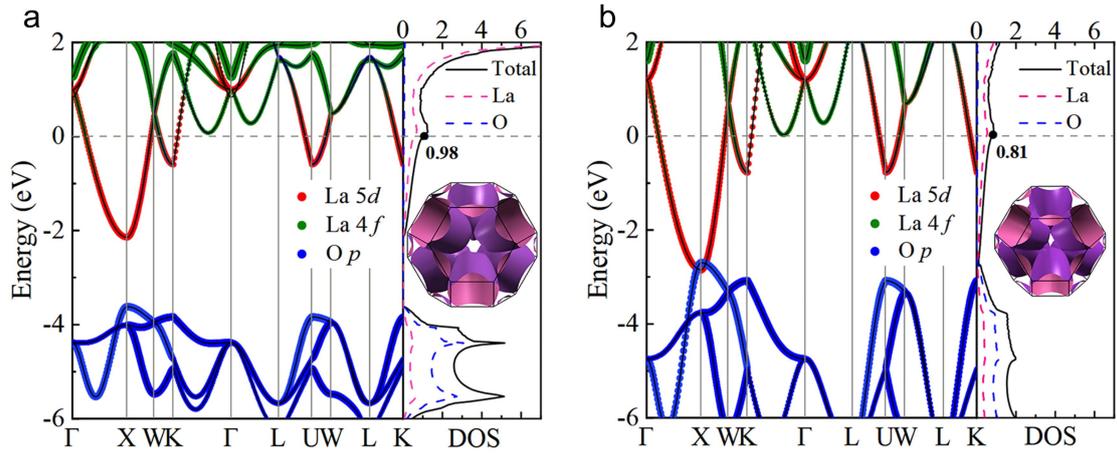

**Figure 3.** Electronic band structures and density of states of bulk LaO in **a.** unstrained and **b.** 8% isotropic compressive strain, with orbital characteristics superimposed on the band structure. The gray dashed line represents the Fermi level. Insets show the corresponding Fermi surface topologies.

**Table 1.** Tight-binding parameters of bulk LaO under unstrained and 8% compressive strain. $t^{d_m}_{[i,j]}$ and $t^{d_m\text{-}d_n}_{[i,j]}$ denote intra- and inter-orbital hopping, respectively, between sites [0,0] and [$i,j$]. All values are in eV.

| Strain | $i$ | $j$ | $t^{d_{x^2-y^2}}_{[i,j]}$ | $t^{d_{z^2}}_{[i,j]}$ | $t^{d_{xz}}_{[i,j]}$ | $t^{d_{yz}}_{[i,j]}$ | $t^{d_{xy}}_{[i,j]}$ |
|---|---|---|---|---|---|---|---|
| 0 | 0 | 0 | 4.4151 | 4.3813 | 2.5289 | 2.5404 | 2.5287 |
|   | 1 | 0 | -0.2099 | 0.0414 | -0.1133 | 0.2514 | -0.9288 |
| -8 | 0 | 0 | 6.4138 | 6.2782 | 3.6365 | 3.6453 | 3.6463 |
|    | 1 | 0 | -0.2011 | -0.0891 | -0.2388 | 0.3179 | -1.2815 |

| Strain | $i$ | $j$ | $t^{d_{x^2-y^2}\text{-}d_{px}}_{[i,j]}$ | $t^{d_{z^2}\text{-}d_{px}}_{[i,j]}$ | $t^{d_{xz}\text{-}d_{pz}}_{[i,j]}$ | $t^{d_{yz}\text{-}d_{py}}_{[i,j]}$ | $t^{d_{xy}\text{-}d_{py}}_{[i,j]}$ |
|---|---|---|---|---|---|---|---|
| 0 | 1 | 0 | -1.497205 | 0.8303 | 0.7068 | -0.0184 | 0.7037 |
| -8 | 1 | 0 | -2.210143 | 1.3439 | 0.7969 | 0.0004 | 0.7924 |

For ToC Graphic Only

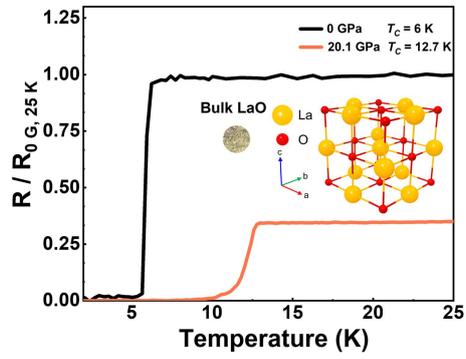

**Supplementary information**

**Specific heat**

The bulk superconductivity is further confirmed by temperature dependence of heat capacity. In contrast to previous studies on thin film LaO, where superconductivity is suppressed with 3 T out-of-plane and 5 T in-plane, our bulk samples require a field of 4 T to fully suppress superconductivity. The difference of heat capacity between 0 and 4 T for LaO is plotted in Supplementary Figure 2, showing a jump at the superconducting transition around 6 K. The onset of superconductivity is 6.3 K and consistent with the $R(T)$ result. We fitted the normal-state (7 - 10 K) heat capacity under 0 T with the equation $Cp = \gamma T + \beta T^3$, and we obtained the Sommerfeld electronic $\gamma$ = 2.3 mJ/mol/K$^2$ and the phonon specific heat coefficient $\beta$ = 0.135 mJ/mol/K$^4$, which corresponds to Debye temperature $\theta_D$ = 305 K. Our $\theta_D$ is clearly higher than that of LaO thin film (262 K) which was evaluated by $\rho(T)$ fitting. As shown in Supplementary Table 2, from $x$ = Te to O, the decrease of $\gamma$ is consistent with the increase of electronegativity while the increase of $\theta_D$ is also expected, owing to decreasing molar mess and atomic radius of $X$. The electron–phonon coupling constant ($\lambda_{e-p}$) can be evaluated by the McMillan equation as 0.68 [15], higher than that of thin film (0.65) and those of La$X$.

**Supplementary Table S1.** Comparison of the lattice parameters and physical properties of *RE*Os in thin films

| Compound | Conductivity | Physical Properties | *a* (Å) | Ref. |
| --- | --- | --- | --- | --- |
| YO | Semiconductor | - | 4.878 | 57 |
| LaO | Metal | Superconductive $T_C$ = 5 K | 5.198 | 15 |
| CeO | Metal | Paramagnetic | 5.089 | 58 |
| PrO | Metal | Ferromagnetic $T_C$ = 28 K | 5.031 | 59 |
| NdO | Metal | Ferromagnetic $T_C$ = 19 K | 4.994 | 11 |
| SmO | Metal | Paramagnetic | 4.943 | 38 |
| EuO | Semiconductor | Ferromagnetic $T_C$ = 69 K | 5.144 | 60 |
| GdO | Semiconductor | Ferromagnetic $T_C$ = 276 K | 4.980 | 61 |
| TbO | Semiconductor | Ferromagnetic $T_C$ = 231 K | - | 62 |
| HoO | Semiconductor | Ferromagnetic $T_C$ = 130 K | 4.904 | 63 |
| YbO | Semiconductor | - | 4.877 | 64 |
| LuO | Semiconductor | - | 4.760 | 65 |

**Supplementary Table S2.** The evolution trend of specific heat in La$X$ system

|  | LaO (film) | LaO (bulk) | LaS (bulk) | LaSe (bulk) | LaTe (bulk) |
|---|---|---|---|---|---|
| $\gamma$ | - | 2.3 | 2.63 | 2.94 | 3.30 |
| $\theta_D$ | 262 | 305 | 276 | 231 | 175 |
| $\lambda_{e-p}$ | 0.65 | 0.68 | 0.52 | 0.42 | 0.34 |

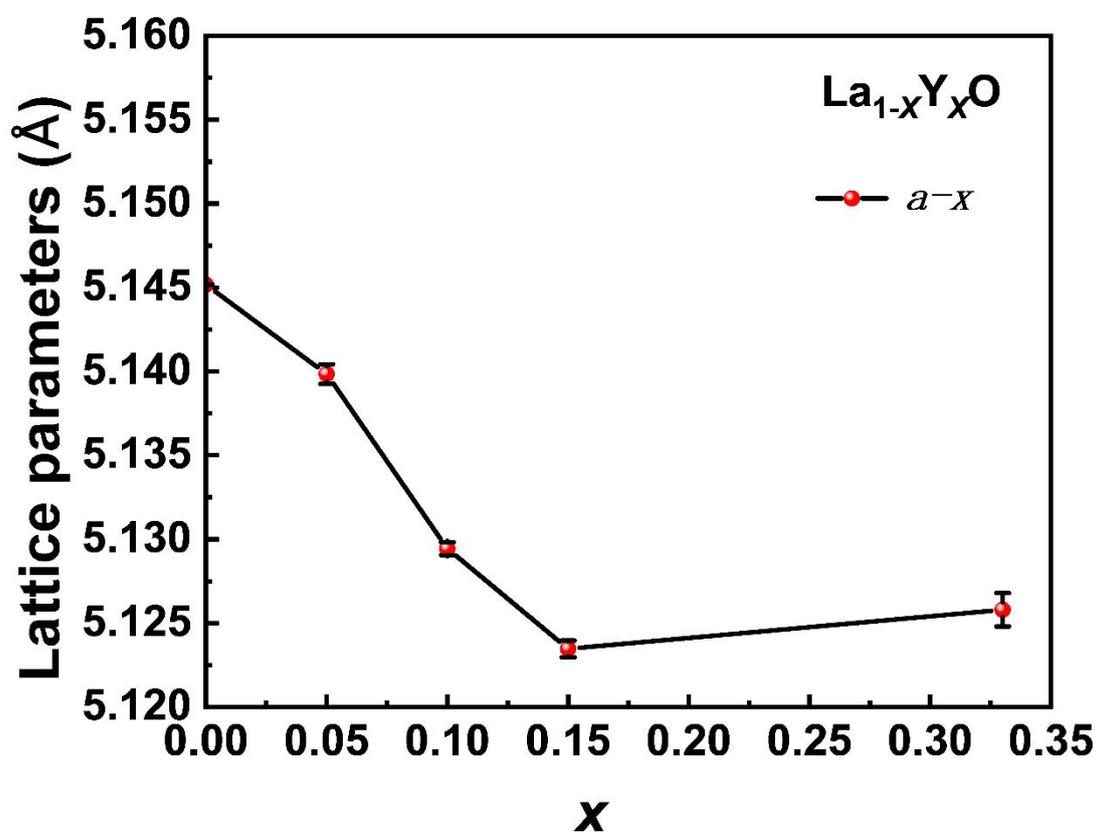

**Supplementary Fig. S1** The trend of the lattice parameter evolution to $x$ in of $La_{1-x}Y_xO$.

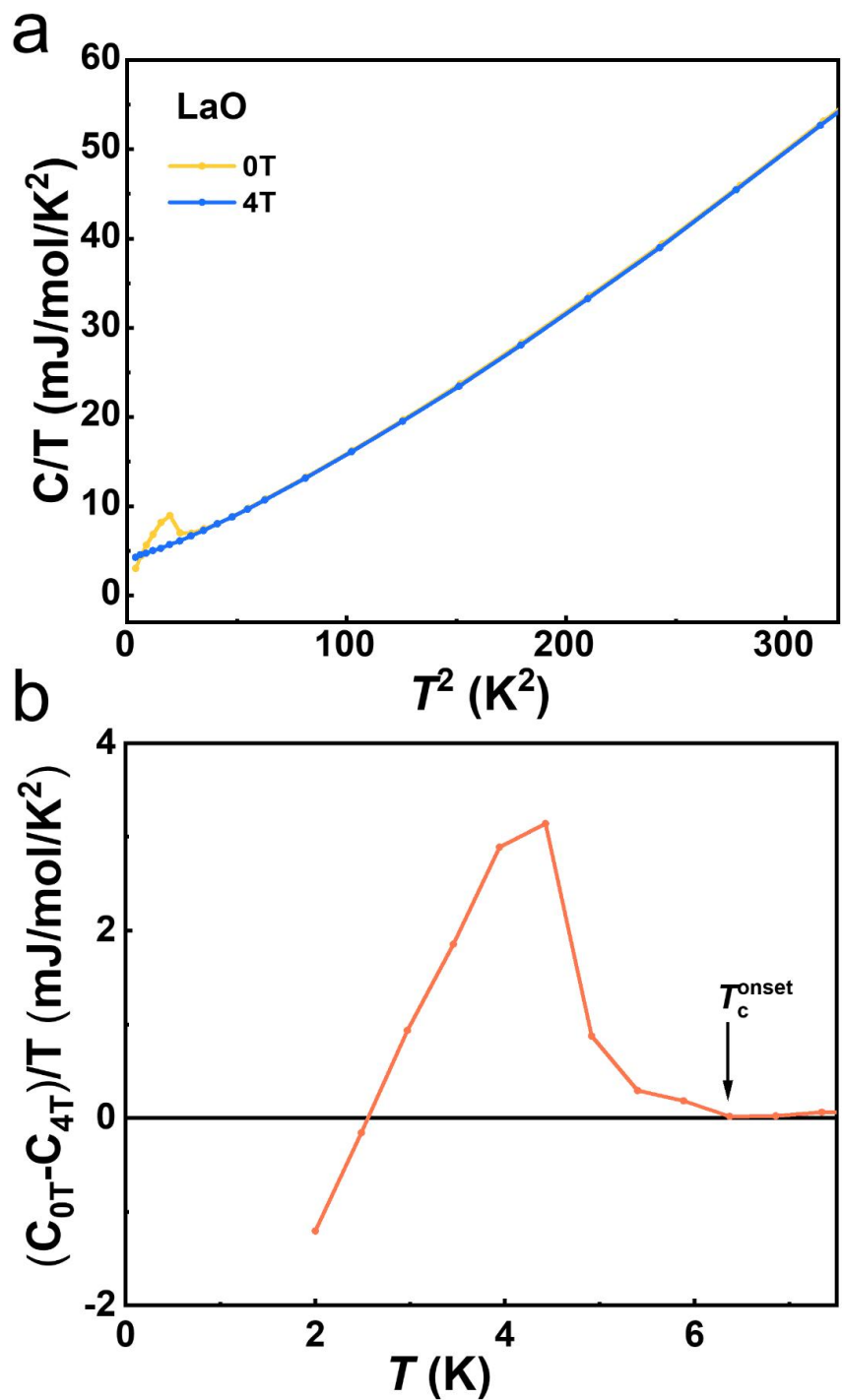

**Supplementary Fig. S2** The heat capacity difference between 0 and 4 T of (**a**) LaO and (**b**) La$_{0.9}$Y$_{0.1}$O

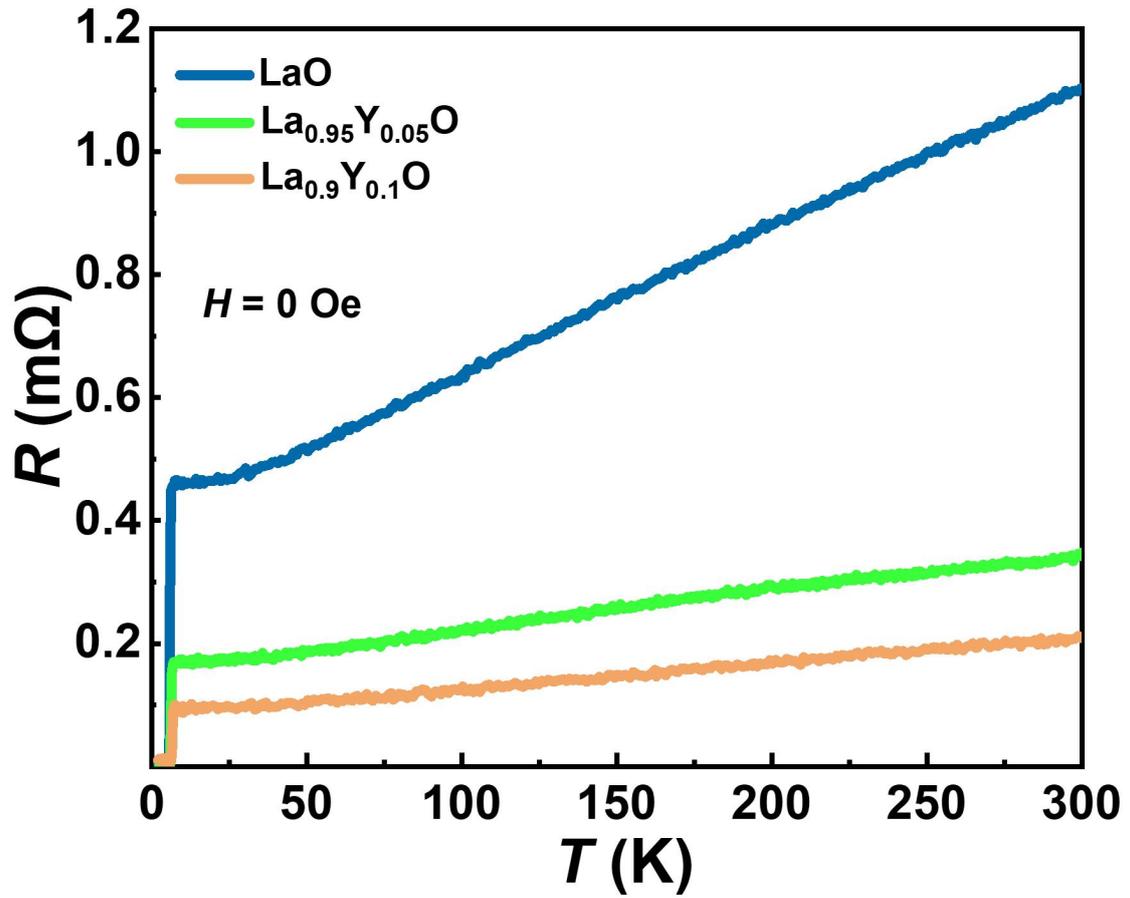

**Supplementary Fig. S3** The evolution of superconducting transition temperature ($T_C$) in La$_{1-x}$Y$_x$O ($x$ = 0, 0.05, 0.10) under magnetic fields (0-2 T).

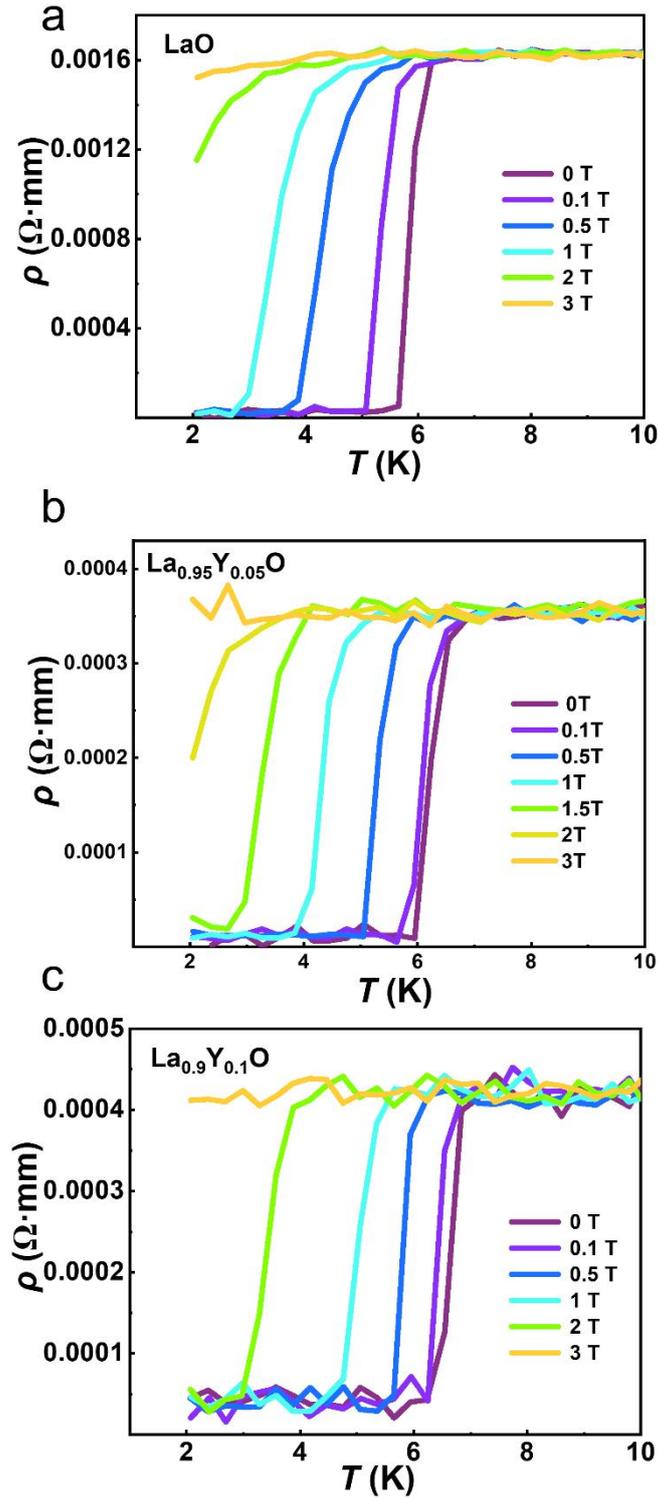

**Supplementary Fig. S4 a-c.** Temperature dependence of resistivity ($\rho$) for **(a)** LaO; **(b)** La$_{0.95}$Y$_{0.05}$O, **(c)** La$_{0.9}$Y$_{0.1}$O, measured under various applied magnetic fields from 0 to 3 T, showing the suppression of the superconducting transition with increasing field.

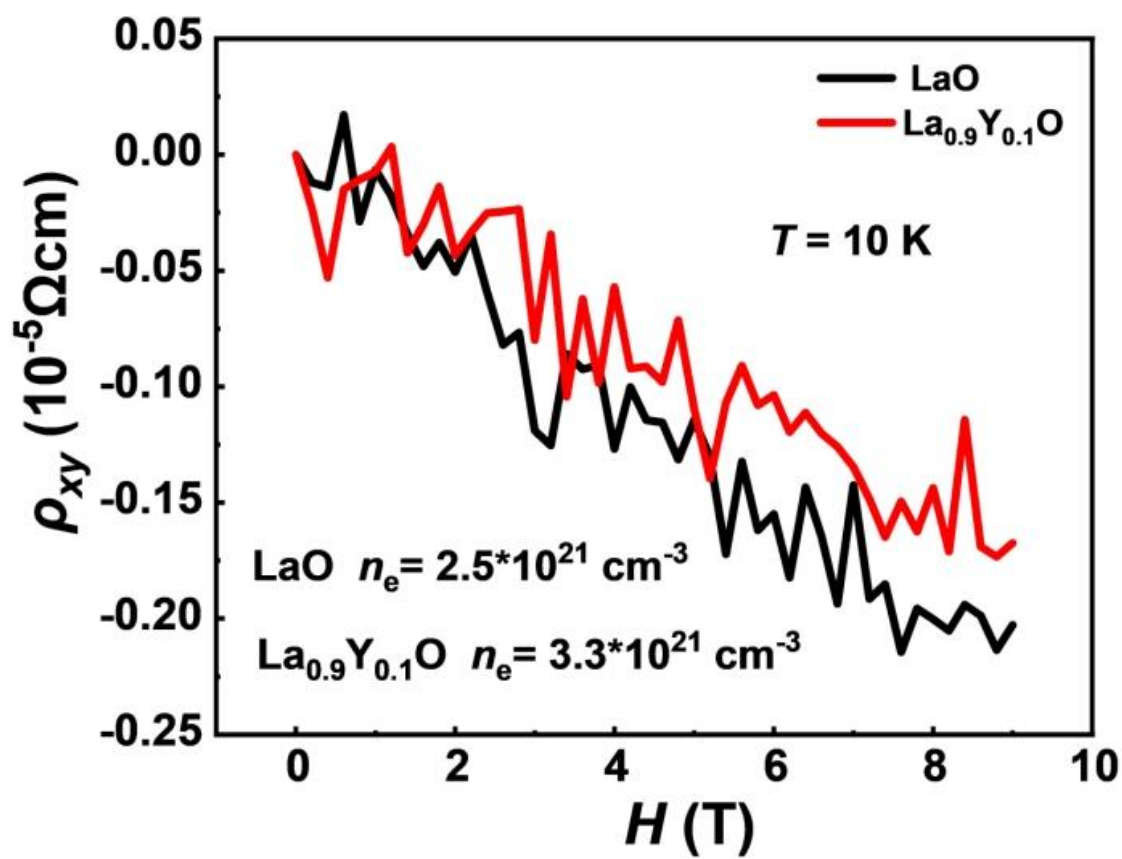

**Supplementary Fig. S5** The Hall resistive response of LaO and La$_{0.9}$Y$_{0.1}$O.

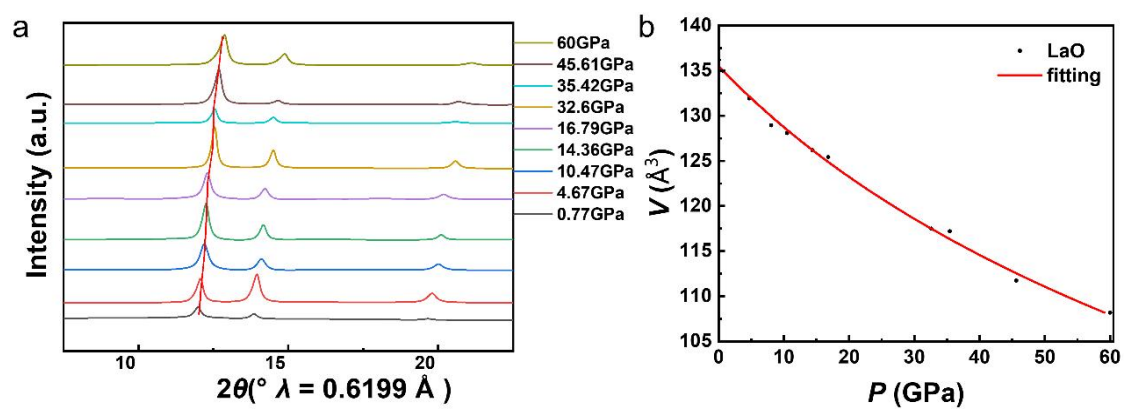

**Supplementary Fig. S6** Structural evolution of LaO under high pressure. (a) In-situ XRD patterns at selected pressure points. (b) Unit-cell parameter versus pressure, showing the experimental data (black points) and the fitting (red curve).

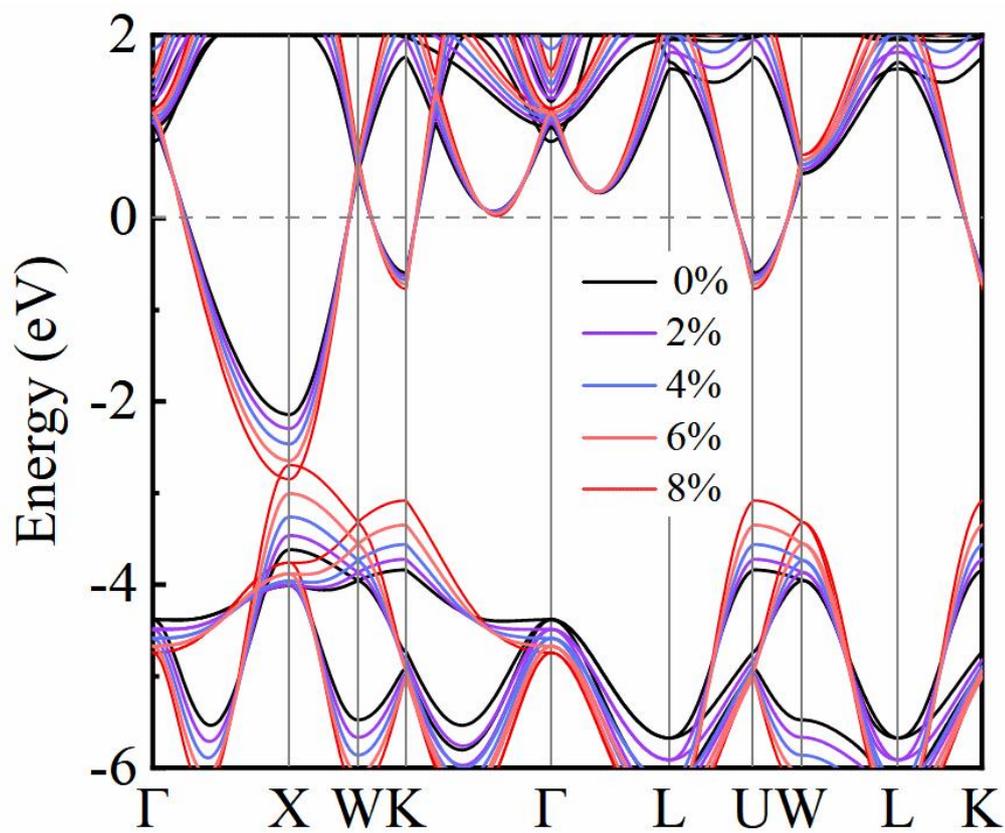

**Supplementary Fig. S7** Calculated band structures of bulk LaO Under 0%-8% compressive biaxial strain.

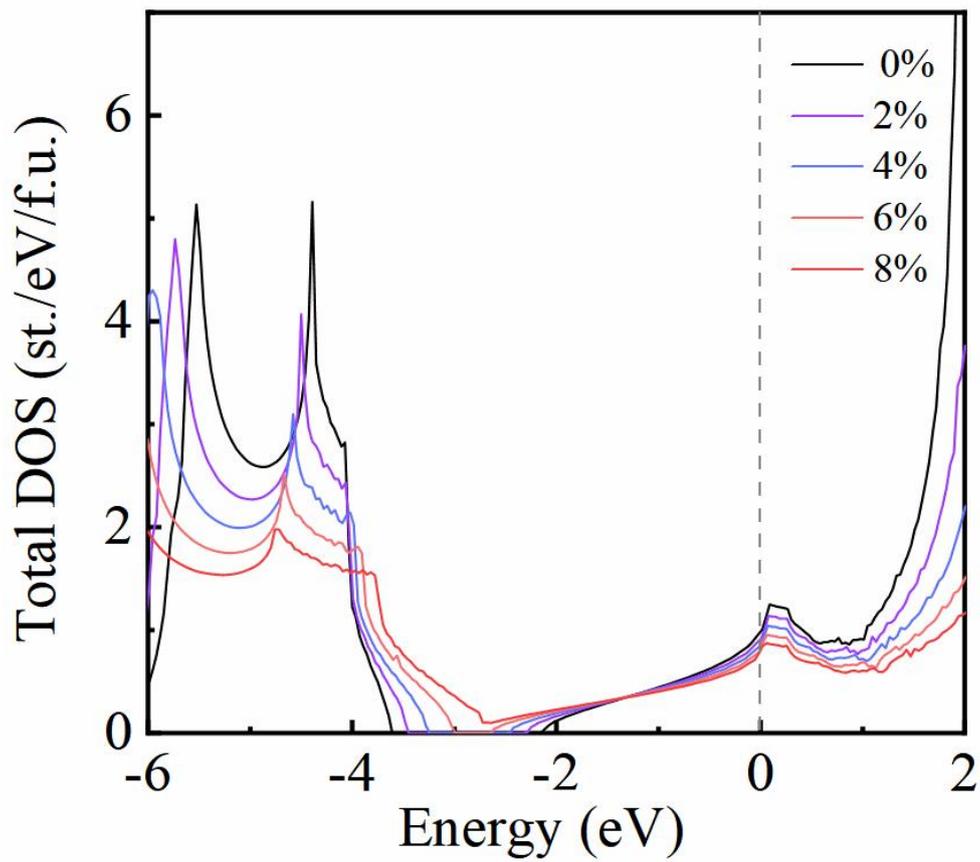

**Supplementary Fig. S8** Calculated density of states of bulk LaO Under 0-8% compressive biaxial strain.

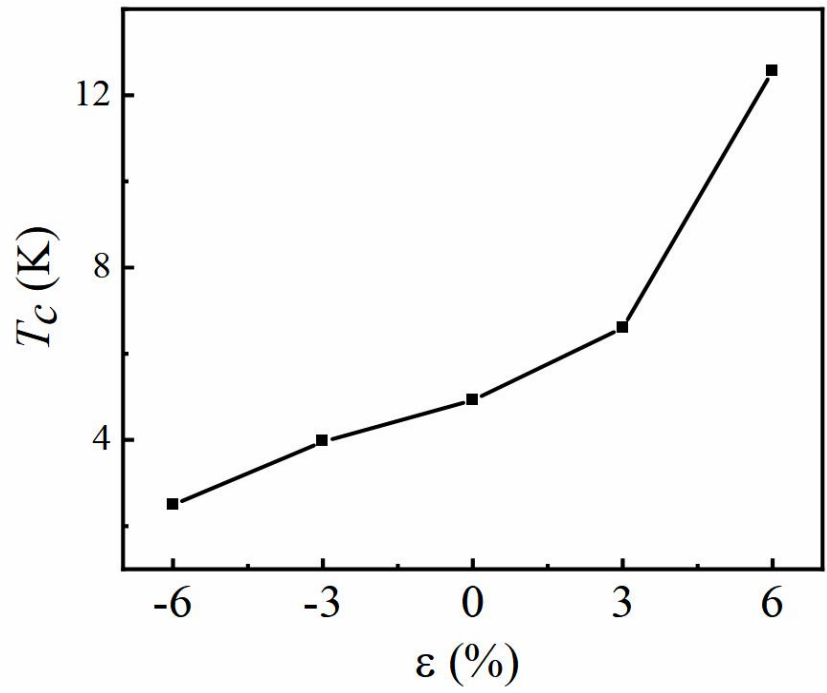

**Supplementary Fig. S9** The superconducting critical temperature $T_c$ of bulk LaO as a function of biaxial strain.

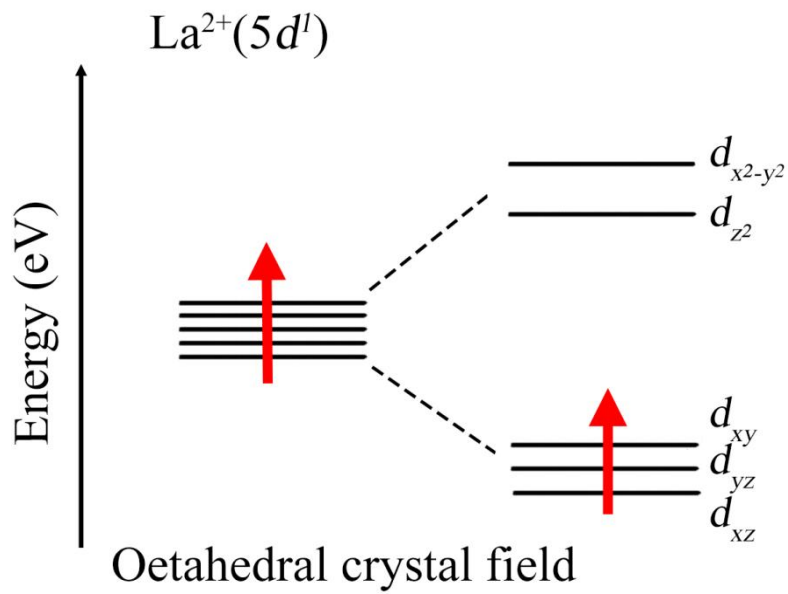

**Supplementary Fig. S10** Electronic configuration of LaO. The arrows indicate nominal configuration $5d^1$.